**Interactive digital storytelling: bringing cultural heritage in a classroom**

Selma Rizvic, Dusanka Boskovic, Vensada Okanovic, Sanda Sljivo, Merima Zukic

University of Sarajevo, Sarajevo, Bosnia and Herzegovina

Abstract:

Interactive digital storytelling becomes a popular choice for information presentation in many fields. Its application spans from media industry and business information visualization, through digital cultural heritage, serious games, education, to contemporary theater and visual arts. The benefits of this form of multimedia presentation in education are generally recognized, and several studies exploring and supporting the opinion are conducted. In addition to discussing the benefits, we wanted to address the challenges in introducing interactive digital storytelling and serious games in the classroom. The challenge of inherent ambiguity of edutainment, due to opposing features of education and entertainment is augmented with different viewpoints of multidisciplinary team members. We specifically address the opposing views on artistic liberty, at one side, and technical constraints and historic facts, on the other side. In this paper we present the first findings related to these questions and to initiate furthering discussions in this area.

I. INTRODUCTION

In Internet and social networks age the perception of information has been changed. The users are bombarded by bits and pieces of information every second. They have no patience any more to read long text or to watch a long video. The time they are ready to dedicate to one topic is shrinking rapidly.

Therefore information presentations methods are also changing. The content is divided in hyperlinked structures, giving the users on the first sight an overview of what they can find inside. Digital storytelling is following this methodology. Nowadays there are online applications presenting the content as a serial of short stories, enabling the viewer to choose how deeply he/she wishes to dive into the information. Still, there is no universal solution for all possible presentation areas.

Handler Miller defines digital storytelling as the use of digital media platforms and interactivity for narrative purposes, either for fictional or for non-fiction stories (Miller 2004). Interactive digital storytelling (IDS) enables the user to influence the flow and sometimes even the content of the story. This new form of conveying information involves professionals from multiple disciplines.

Although the area of use for interactive digital storytelling is expanding fast, in our research we focus on its application in virtual cultural heritage presentation and serious games. Our experience in this area led to guidelines for IDS presentations of cultural heritage, presenting how documentary historic information about a heritage object or site can be conveyed in an engaging and immersive way to general audience in museums and online.

In our attempt to develop a new method for interactive digital storytelling: hyper-storytelling, we engaged experts from computer science, visual arts, film directing, literature, psychology, communicology and human computer interaction. They have analyzed a sample interactive digital storytelling application and offered their insights and recommendations to be embedded in the new methodology.

In order to bring the cultural heritage in the classrooms and employ multimedia applications, especially IDS and serious games we need to extend our team and include pedagogy expertise. The primary objective for introducing these modalities of content presentation is motivation for learners, and accommodating the preferences and attitude of millennials. This could be summarized into one important feature: edutainment. Edutainment is not a novel attribute in designing educational applications, but the importance of edutainment is especially recognized in the research field of serious games and digital cultural heritage presentations (Philpin-Briscoe et al., 2017). It is notable that serious games can create more engaging educational practices leading to 'the emergence of serious games as a new form for education and training'(DeFreitas and Liarokapis, 2011)

Difficulties in discussing and measuring edutainment arise from opposition between the important pedagogical aspects and aspects of importance for the entertainment part of the content presentation as identified in (Wiberg and Jegers, 2003), identifying necessity of tradeoffs in balancing both the entertainment and the educational part of the application.

In this paper, based on our experience, we present the guidelines upon which the hyper-storytelling method will be founded, but also in addition we address challenges of successful introduction of the IDS into classrooms. In Section 2 we offer a brief overview of related work in IDS field with particular emphasis on virtual cultural heritage applications and serious games. Section 3 presents lessons learned from evaluations of the sample IDS application. The part of the evaluation process was conducted in order to introduce the interdisciplinary team members with the topic and goals of the research. The Section 4 summarize the recommendations of interdisciplinary experts in form of guidelines for future IDS applications. Section 5 presents results of the user evaluation survey designed to explore the challenges of balancing the education and entertainment, and also opposing views on artistic liberty, at one side, and technical constraints and historic facts, on the other side. We conclude with the future work directions.

II. RELATED WORK

California State University, Chico, class generated a five part definition of digital stories, according to which, for assessment purposes, they should: include a compelling narration of a story; provide a meaningful context for understanding the story being told; use images to capture and/or expand upon emotions found in the narrative; employ music and other sound effects to reinforce ideas; invite thoughtful reflection from their audience(s) (Alexander, 2011). This definition introduces some key words for our research: narration, images, music, emotions. It shows that only multidisciplinary teams can combine all these notions into an interactive application.

All authors in the literature agree that the foundation for successful IDS applications is the skilful use of general storytelling principles defined through history in all kinds of media. Aristotle's seven golden rules: plot, character, theme, dialog, music, decor, and spectacle, are easily recognized in engaging and immersive interactive digital stories.

Hero's journey is another storytelling structure which is or could be used in IDS. It is a pattern of narrative identified by the American scholar Joseph Campbell that appears in drama, storytelling, myth, religious ritual, and psychological development (1949). It describes the typical adventure of the archetype known as The Hero, the person who goes out and achieves great deeds on behalf of the group, tribe, or civilization. The proposed structure consists of 12 stages, starting with introduction of Hero's world, describing the call for adventure and following him through different obstacles until the desired goal is fulfilled. This structure is better suited for adventure movies and novels then for documentary narrative, but some elements could be applied.

As the focus of our research is IDS in virtual cultural heritage applications and serious games, we present here some examples of such projects and discuss their advantages and drawbacks.

Etruscanning 3D project (Pietroni et al, 2013) is an IDS application created to present findings from Etruscan Regolini-Galassi tomb through an interesting combination of storytelling with 3D environment of the tomb and interactive models of artefacts found there. The user stands in front of the screen where the virtual environment of the tomb is projected and interacts with the application using gesture-based interaction interface with Kinect motion sensor. This project is a great example how to use storytelling to convey to the user the history and importance of archaeological findings. However, gesture-based interface limits its usability to a museum setup.

Admotum application created within the Keys to Rome exhibition on Roman culture during the rule of Emperor Augustus, is a serious game engaging the user in the quest for objects from 4 involved museums (Pagano et al, 2015). The exhibition was held in 2014 at the same time in Rome, Alexandria, Amsterdam and Sarajevo, with goal to present different parts of Roman Empire at that time through a combination of museum collections and digital content. Admotum application was designed as a treasure hunt, where the users explore first the virtual environments of Roman objects from their location, and, upon finding all objects, they can unlock virtual environments from remaining three locations and look for their objects. Storytelling plays significant role in this application, as every virtual reconstruction and particular objects are described through narrations of virtual characters. However, the users are so engaged with not simple gesture-based navigation, that most of them do not pay much attention to stories they hear.

The mentioned projects open new research questions which will be addressed within our research for a new IDS methodology. The first question is how to obtain maximum user immersion in the virtual presentation of cultural heritage. Secondly, it is important to build applications with high level of edutainment, e.g. to convey enough information to educate the user in historical context of heritage object or site, while making him/her amused and engaged by the presentation. Virtual environments with interactive storytelling enable the user to watch stories on demand. This advantage could turn into a drawback if users do not watch all offered content. In literature this is called solving the narrative paradox. It is the conflict between pre-authored narrative structures, especially plot, and the freedom a virtual environment (VE) offers a user in physical movement and interaction, integral to a feeling of physical presence and immersion (Schoenau-Fog, 2015). The new IDS method should introduce a motivation factor for users to view the whole offered content.

### III. Lessons learned from evaluation of IDS application

In order to facilitate the work of interdisciplinary experts on development of a new IDS method hyper-storytelling, there was a need to familiarize them with typical demands of IDS applications for cultural heritage presentations and serious games. Such applications usually consist of stories, interactive 3D models of cultural heritage (CH) artifacts and interactive virtual environments (IVE) presenting reconstructions of cultural monuments' original appearance. The users can virtually explore the IVEs, watch or listen to the stories and learn about the purpose and historical context of selected objects. The applications are usually on-line or accessible for mobile download, but they can as well be set up in a museum. Some of them introduce augmented reality elements for combining the digital content with landmarks and on site elements.

User experience evaluation studies of existing IDS applications have shown the following major drawbacks:
- stories are too long to keep the attention of users
- users have problems with navigation in IVEs and do not find triggers for all stories
- the content is missing the motivational factor which would keep the user engaged until all of it is explored
- there is too much information which is not well structured and makes users bored
- the application does not give satisfactory user experience to all audience target groups
- serious games for cultural heritage are too easy or too difficult for playing

*A. The sample IDS application - White Bastion*

White Bastion is a fortress overlooking the city of Sarajevo. It has been guarding the access to the city since medieval period. Through its history it has changed appearance several times.

4D virtual presentation project (Rizvic et al, 2016) introduces Internet users with the history of the fortress, its appearance from medieval, Ottoman and Austria-Hungarian period till present times, through interactive digital stories. This application has been selected as a sample IDS application for introducing the interdisciplinary team of experts, and their experience developed into the guidelines for interactive digital storytelling presentations of cultural heritage (Rizvic et al, 2017). The following paragraphs summarize the guidelines.

The White Bastion application structure is shown in Fig.1. The user can watch on demand 10 digital stories and explore 6 interactive virtual models of the fortress. The intro story presents the overview of the content implemented inside the application. Stories about medieval, Ottoman and Austrian-Hungarian period offer more details on events and characters related with the fortress in those times. Some of interactive virtual environments also contain stories about particular parts of the fortress and its inhabitants. They also contain models of digitized archaeological findings from the site with their virtual reconstructions. This application can easily be set up in the museum, possibly next to the artifacts found on the site, which will just add to the visitors' experience and not influence their perception of the application itself, nor the proposed guidelines.

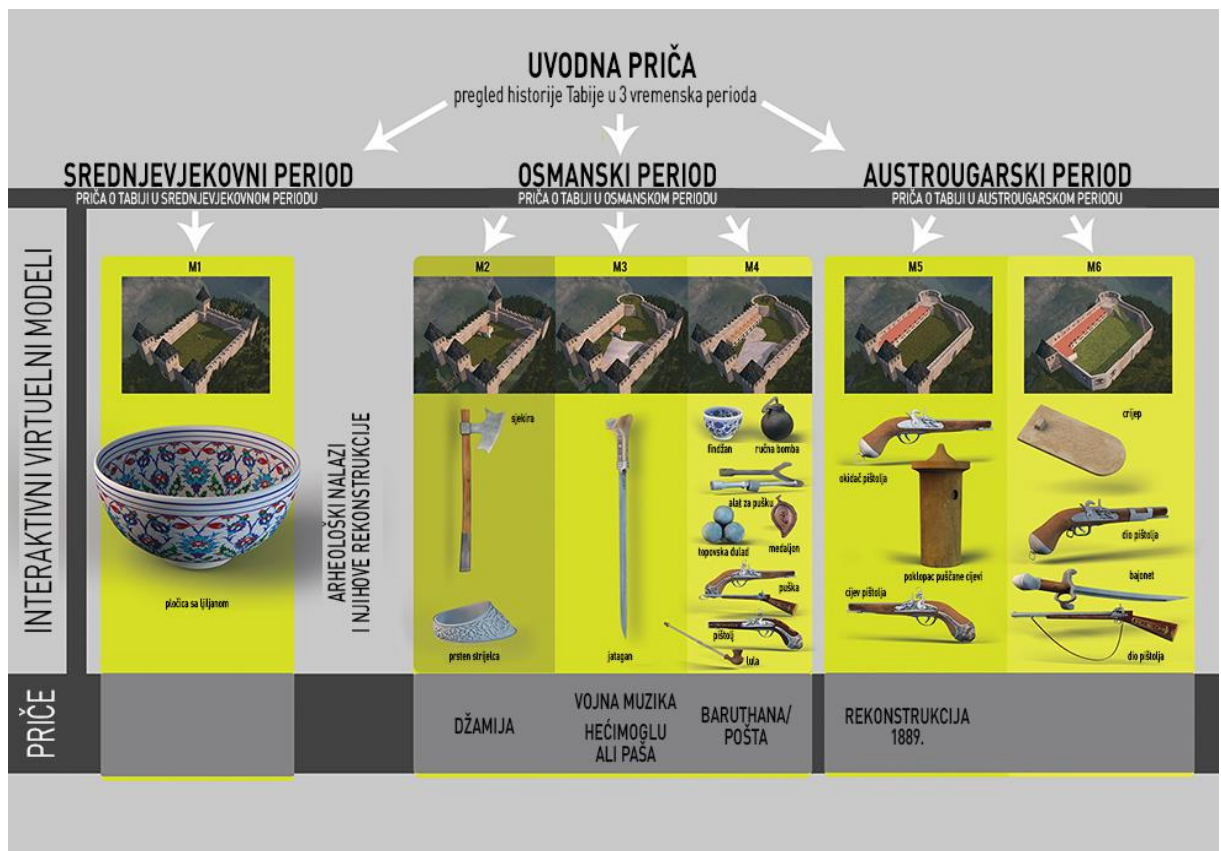

**Fig. 1** White Bastion project structure

Initial user evaluation study has shown some drawbacks in the concept of White Bastion IDS application. Most of the users have not seen all digital stories. Some of them could not orientate in interactive virtual environments and find stories inside them. Although majority of participants have appreciated presented IDS concept, particularly the narrator character, the eternal soldier of the fortress, they drew our attention to possible improvements. As the initial user evaluation has been conducted with variety of user target groups (Bosnian, non-Bosnian, different computer literacy, different ages and professional backgrounds), we considered as useful to evaluate the same application with interdisciplinary experts. This evaluation will identify the flaws of the concept and contribute to new IDS method development.

*B. The interdisciplinary team*

A common issue in new on-line media research is that it is constrained to just one scientific field. Interactive digital storytelling, as a novel form of media communication, needs to be considered by experts from all relevant fields in order to develop a new methodology appreciated by all user target groups. Here we present arguments for engaging particular multidisciplinary experts in our research team.

   1. Computer science

Computer science is a point of synthesis for all team members' contributions. The computer scientists will implement the new IDS method on Internet media. In the method development phase they gather information and recommendations from remaining team members and decide with them which of their recommendations should be incorporated in the final research product.

   2. Psychology

One of the main goals of the new IDS method is to transfer knowledge to the user. Cognitive psychology offers a large set of tools to be implemented in this research to obtain the highest level of learning the presented information. The hyper-storytelling method will be developed and evaluated following suggestions of this expert who will identify its advantages over existing IDS methods.

   3. Communicology

Internet media has its rules and laws of presenting information. The communicologist contributes to this research by suggesting the most appropriate ways of adjusting the IDS to this media, with goal of obtaining maximum user immersion into the story/application. The new IDS method should produce educational, fun and easy to use IDS applications. The communicologist will offer patterns for achieving this goal in the novel way.

4. Film arts - directing

Interactive digital stories created using the new method contain a large amount of film forms. In fact, they are a combination of movies and interactive virtual environments such as in computer games. The director has a leading role in creation of these movies, as well as directing the interaction in the VEs, to obtain the most user friendly combination and enable the user to learn the presented information.

5. Literature - storytelling

The writer of interactive digital stories scenarios is adjusting his storytelling methodology to the Internet media. Now he has on his disposal an enhanced set of tools, instead of pure text, to tell the story. In collaboration with the director and other team members, he decides on narrative method, characters and user interaction, in order to raise the attractiveness for the users.

6. Visual arts - graphics design

Graphics design expert is in charge of the visual appearance of the IDS application, starting with user interface, until the picture elements distribution and shot composition. He needs to define a set of rules to be applied on the IDS application to make it visually appealing for the user, who should interact with it in an easy and natural way

7. Human-computer interaction (HCI)

For a long time the success of products which involve interaction has been measured by the quality of user experience. There are different existing methods and approaches for user experience evaluation, some of them general, some linked to a specific application domain. The expert from this field will develop and apply a new method for user experience evaluation dedicated to the IDS. Through a set of user studies we will assess the advantages of the proposed methods. The evaluation will encompass the quality of interaction and information perception, and focus on the following heuristics important for digital storytelling: user immersion and edutainment value.

*C. The user experience evaluation study*

We conducted several user experience evaluation studies of the White Bastion application. The initial user experience evaluation results (Rizvic et al, 2016) have shown that users appreciated interactive storytelling more than linear one, empathized with narrator character and learned information about this cultural monument in an attractive and immersive way. The main drawback of the application was, according to them, the difficult navigation in IVEs which prevented them to find triggers to all digital stories.

*The psychologist* has evaluated the White Bastion application from the aspect of cognitive load theory (Sweller et al, 2011), integration of cognitive load theory and concepts of human computer interaction (Hollender, 2010) and process model of hypertext reading (DeStefano, 2007) applied to hypermedia. He divided his observations into macro and micro levels. Appreciating the home page as an "organizer" which would decrease the cognitive load as recommended also in previous evaluations (Chalmers, 2003), he noted the possibility of esthetic and functional improvements to make the page more dynamic and attractive. The possibility for users to choose their way through the presented information was recognized as respecting the segmentation principle which contributes to optimization of the intrinsic load. Rather long loading time of interactive virtual environments he qualifies as decrease of the presentation's usability. On the level of digital stories, he appreciated the possibility for users to control the presentation. Finally, he noted redundancy in information presentation in occasions when the same information is presented visually and also in the narration simultaneously. These concurrent presentations compete and increase cognitive load.

*The communicologist* stated that in the American modern journalism there are two kinds of genres: news and story. According to (Quintillian, 2006), in order to be communicable, the presentation of information needs to contain three elements: to have logical foundation, to be attractive and to be ethical. She found that White Bastion, as a novel way of presenting information, invokes curiosity founded on known denotations and motivates users to stay with it till the end. The presentation could be enriched with adding legends or personal experiences of the narrator-soldier character, such as love, prayer or feelings. She recommends addition of sign language interface for persons with special needs.

*The graphics designer* has noted the lack of unique visual style for user interface of the web site and design of digital stories' elements. The whole presentation should have had a logo, coherent font selection, and navigation and screen elements. He appreciated the esthetics of video production. An intro sentence or short sequence should have been added to introduce users what they could expect from the presented content. Virtual environments could have been more realistic with better illumination and render quality.

*The storywriter* states that the main quality of interactive information presentation in White Bastion application is entering the creative field of possibilities, which enables us to follow the already established reflexes of the recipient who is an experienced consumer of cyber contents and interactive communication and offer him/her to gain knowledge in such familiar way. Drawbacks are in not implemented possibility to further use creative tools for pageant of different situations from the story through film or visual arts.

*HCI expert* has performed the heuristic evaluation of White Bastion application (Boskovic et al, 2017) using instrument based on well-established ten principles of interaction design (Nielsen 1995). She noted that the clickable objects in VEs are not enough emphasized and the navigation in VEs is not intuitive. It would be useful to introduce a Help section where the user could be introduced with most important features of user interface.

*The film director* considers that being the White Bastion implemented with minimal budget is not noticeable on the implementation of the project. The content is interesting, coherent and systematic. He appreciates the digital stories as poetically stylistic and the excellent performance of the actor narrator who was interpreting historical facts in real location. All contributing elements such as text, costume and rhythm should be appreciated. The application succeeded to recreate the atmosphere of the past times in a way attractive to the modern spectator. The general remark is absence of stylistic unity of the "gestalt", particularly a visual one, being it a basis of the media we are dealing with. The visual aspect on the level of information theory is the first step of keeping the consumer, as it creates the decisive first impression.

## IV. Guidelines for Interactive Digital Storytelling

As Denard established *the London charter* (2012), internationally-recognized principles for the use of computer based visualization by researchers, educators and cultural heritage organizations, we present a set of guidelines for interactive digital storytelling presentations of cultural heritage. These guidelines are provided by interdisciplinary experts from computer science, visual arts, literature, film directing, psychology, social sciences and human computer interaction after their evaluation of a sample IDS application White Bastion.

*The psychologist* concludes that for IDS to fulfill the usability criteria, we need to decrease unnecessary cognitive load, adjust the intrinsic and increase the relevant load. According to learning principles derived from cognitive load theory, a digital story will be more educationally efficient if:
- both text and pictures are used, instead just text (multimedia principle)
- text and pictures are temporary integrated (principle of divided attention)
- graphics content is used with narration instead of written text (modality principle)
- one information is not presented in two or more forms (redundancy principle)
- key information is emphasized within information organization structure (signalization principle)
- unnecessary material is excluded (coherence principle)
- material is presented in segments and the user has control over them, instead of linear structure (segmentation principle)
- for beginner users all kinds of organizers are provided
- the application is easy to use

*The communicologist* has a different opinion regarding redundancy. She states that qualitative redundancy should be introduced in IDS. She recommends the hourglass narration structure for storytelling. The information should be segmented in short structural units, connected similarly as 1001 night stories, constantly keeping the user's attention, even attracting him to come back to the story. Key words for the new IDS concept should be attractiveness and edutainment value. Apart from the narrative, very important are richness of picture (colors) and sound, significantly contributing to attractiveness of the application.

*The graphics designer* emphasizes importance of user experience quality. The user interface should contain movement and animation to enrich the overall satisfaction. Design trends and technologies should be followed and applied. Lately parallax scrolling has showed great results for interactive storytelling applications. The users appreciate to be lead through presentation without investing much effort, but we need to leave a detail or two for those who like exploring. Historical storytelling should be enriched with a human element (for example hand drawing) to create some warmness and soften the communication. Unique visual style of all used elements (web site, individual stories, live footage, text and illustrations) enhances user satisfaction.

*The storywriter* has described interactive perspectives of digital storytelling through the following elements:

Storytelling: Art of storytelling follows socio-cultural and techno-aesthetic changes. The digital revolution has opened new possibilities for storytelling techniques. Digital storytelling is becoming creative combination of accessible technologies enhancing the traditional approach through unity of picture, sound and movement with the narrative, in order to transfer the story message.

Multimedia narrative: Multimedia is a result of combination of text, picture, sound, photograph, animation, movie and other types of new media, in cyber space also the hypertext. It is extremely important to arrange the narrative aspect with visual, interactive and compositional elements in a functional entirety. The content should not have unique flow of reading/watching/listening, but the user should be able to determine it dynamically.

Process model: Narrative content is open for interaction and connecting various media elements in one entirety. The most important task in developing this new media and storytelling expression is definition of an efficient process model.

Narration visualization: Design process is never the idea of only one of its structural elements, but a process, a set, a coherent thought about external and internal, a detail as well as a whole. Digital idea design means conception of visual style content components, derived from the demands of narrative. Visual component is nowadays a foundation of human communication, the same and even more than the text.

Interactive narrative: The goal of interactive narrative is not to author the story, but to offer the contextual building blocks and environment where narration could be discovered or built by the user. The key concept of interactive narrative is the ability of users to make decisions on the narrative.

Goals of interactive narrative: Successful interactive narrative is designed according to the user needs to provide him/her a pleasant and inspiring experience. This includes all experience aspects (physical, sensual, cognitive, emotional and esthetic). The first step in creating such narrative is clear definition of the main message and emotion of the story, as often neglected category of narrative aspect.

Structure of interactive narrative: Interactive project starts with developing a concept which connects the user experience and intellectual involvement with the user interface and contents. The story is experienced through its content elements instead as a linear narrative. This includes creation of a structure in which the content components will be arranged to form the complete narrative user experience.

Motivation in interactive narrative: The curiosity of the user is a moving force through interactive story. It motivates him/her to pass all steps of interaction.

Interaction design: User experience could be divided in 3 aspects: the form - dealt with by graphics design through creation of a visual language for communicating the content; behavior - shaping user's behavior towards the story, its form and content; content - created by animators, sound artists, and information architects.

*The HCI expert* states that, based on 10 Nielsen's usability heuristics (1995), IDS applications should provide users with a sense of control and consequently: contain information on navigation in VEs, emphasize trigger objects for certain actions and enhance integration of interactive 3D geometry and narrative content. For successful user evaluation of IDS applications, Nielsen's heuristics should be extended with evaluation of experience related to content (content itself, personalization, strategy, presentation modes interconnection), as well as experience related with navigation through the story and within interactive virtual environments.

*The film director*
Since discovery of the first camera, film art has been founded on tradition of novels, drama and performing arts. After that, the sound was introduced, followed by music. Various elements joined through development of film technology, but film art still remained the same as at the beginning telling a story. Interactive digital storytelling needs to inherit all elements taken over from film, to follow the film language syntax and grammar, to appreciate scenario as a movie on paper, to respect film's internal logic and convey its message. If a movie director was asked the difference between film and theater directing, he would say it is the same except the tools are different, according to the media. This applies to IDS as well. Furthermore, such a novel media brings us an opportunity and a challenge to become pioneers of its poetics.

Particular recommendations for the new IDS method would be the following: stories have to be short, informative and dynamic; structure should be defined by a content editor(s) instead of a director; actors should be used to enrich documentary information and add emotion to it.

*Computer science experts* emphasized importance of measuring immersion and edutainment value of IDS applications, as a quantitative parameter for their evaluation. They state that a crucial element for IDS applications, particularly those which contain interactive virtual environments, is to solve the narrative paradox and motivate users to explore all offered content completing the story. Interactive virtual environments introduce additional immersive aspect to interactive digital stories, as users have possibility to browse recreated cultural monuments and watch related stories inside them. IDS applications need to be portable to all platforms, desktop and mobile devices.

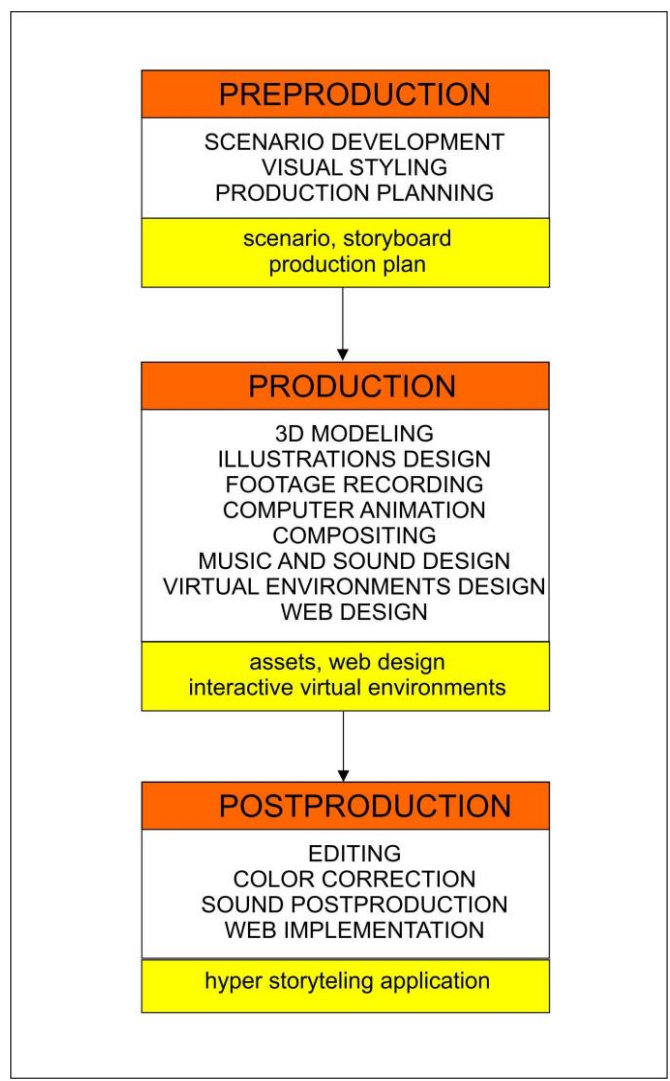

**Fig. 2** IDS workflow diagram

Adding up all presented considerations, we can summarize them in the following guidelines for IDS presentations of cultural heritage:
- engage professionals for all content creation fields
- all content has to have unique visual identity

- use multimedia and virtual reality
- divide content in sub stories which can be watched independently
- stories should be short, dynamic and informative
- use characters to communicate emotion and raise edutainment value
- introduce motivation factor to solve the narrative paradox • create IDS application to be platform independent

According to these guidelines the hyper storytelling IDS application development process can be described by workflow diagram in Figure 2.

The diagram shows that development of hyper-storytelling application consists of preproduction, production and post production stages. This is a common workflow for digital production process (Kerlow, 2009), but we extended it with a number of elements particular for IDS. In preproduction the producer, director and visual artist should agree upon the scenario and visual styling of the application. All production planning activities (actors casting, location scouting, team members selection, budget planning) are performed in this stage. The production will be performed according to the scenario and storyboard, the main results of this stage. Production stage includes all assets creation (music, illustrations, footage, computer animations, 3D models), web design and design of interactive virtual environments. In postproduction stage all results of the previous stage are put together through editing and implemented on the web site.

## V. Exploring the challenges

Working in multidisciplinary team highlighted different opinion on benefits of the multimedia cultural heritage applications, and also priorities in IDS presentations. All team members agreed on evident significance, but opinion on objectives and how to accomplish them were quite diverse. These differences were highlighted in the educational context: either within museum or standard classroom.

Studies evaluating benefits of introduction of cultural heritage IDS in the classrooms researchers usually address efficiency and influence on learners and learning process. The benefits for both aspects has been conformed in the past, summarizing the results of several studies showing the ability for serious games to engage both young and older learners, both experienced gamers and non-gamers, and showing the efficacy of the game (deFreitas & Liarokapis 2011).

Experience of multidisciplinary team motivated us to conduct the user evaluation addressing the standard benefits as motivation, increased understanding and learning and greater efficiency, and disadvantages such as technical problems, learner motivation and readiness (Coomey & Stephenson 2001), but to compare the evaluation results across different professions. We decided for the self-selection of users, with emphasis on their reliability, and openness to convey negative opinions. In order to ensure that the responses represent diverse cross-section of respondents and to establish validity of survey responses, we included questions for relevant demographic data: age, professional background and status (Lazar et al., 2010).The status refers to distinguishing between students and teachers, since these were two the most interesting populations for us.

*Procedure*

The evaluation involved 46 participants. Presentations of multimedia digital heritage is intended for a broad audience, but although we aimed to balanced different user types regarding their professional background (art, engineering, and humanities) we were focusing on students and teachers population.

The users were invited to take a tour within the White Bastion presentation, and/or two additional cultural heritage IDS: Kyrenia and Bridges of Sarajevo. After watching the IDS the users needed to spend additional 10-15 minutes for answering web based post-interaction questionnaire.

*Questionnaire*

The questionnaire used is very simple. After the introductory part with data for user profiling, main part for contained comprises 3 sub-scales addressing importance of IDS for education, features of IDS contributing to success and features of users contributing to success; containing total 9, 8 and 4 Likert items, respectively. Likert scale items were defined as straightforward statements with positive logic, repetitions were avoided.

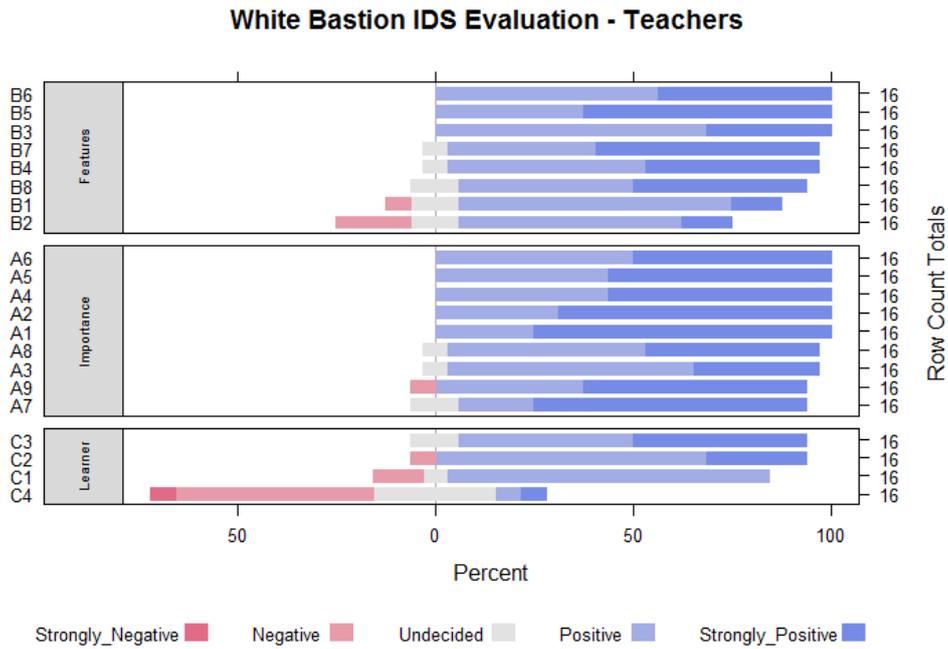

**Fig 3** IDS evaluation – Teachers

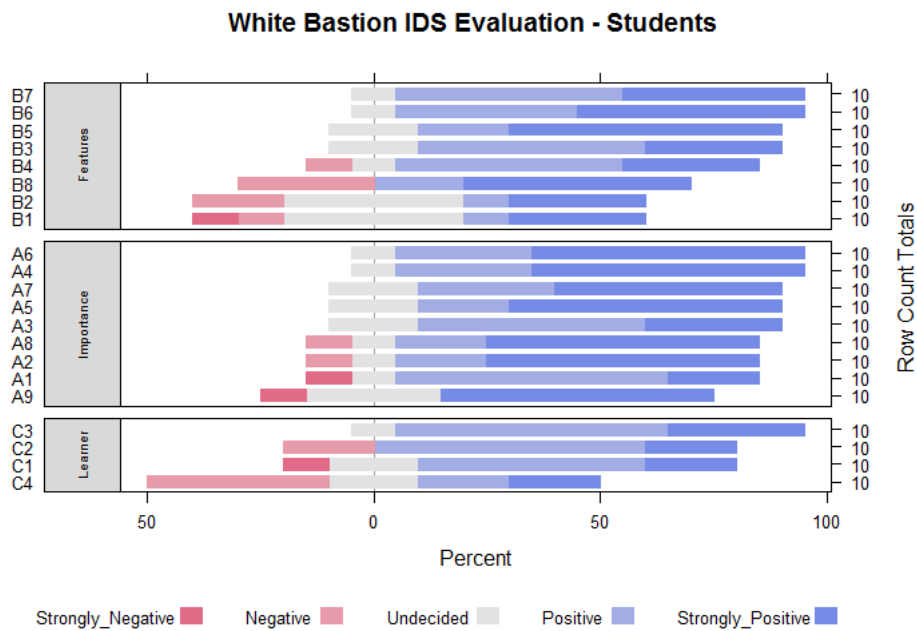

**Fig 4** IDS evaluation - Students

Results are analyzed separately for different user profiles regarding the professional interest (Arts, Humanities, Engineering & Math) and the role in education (Teachers, Students). Visualization provides insight in assessment of answers for each specific item in the main part of the questionnaire, as presented in Fig. 3 to Fig. 8. Distribution of responses for digital stories and interactive digital model are presented in accordance with (Heiberger and Robbins, 2014).

Comparing distribution of answers in Fig. 3 and 4, it is easy to notice that students find less importance in using IDS (Importance part) and also in specific features of IDS (Features).

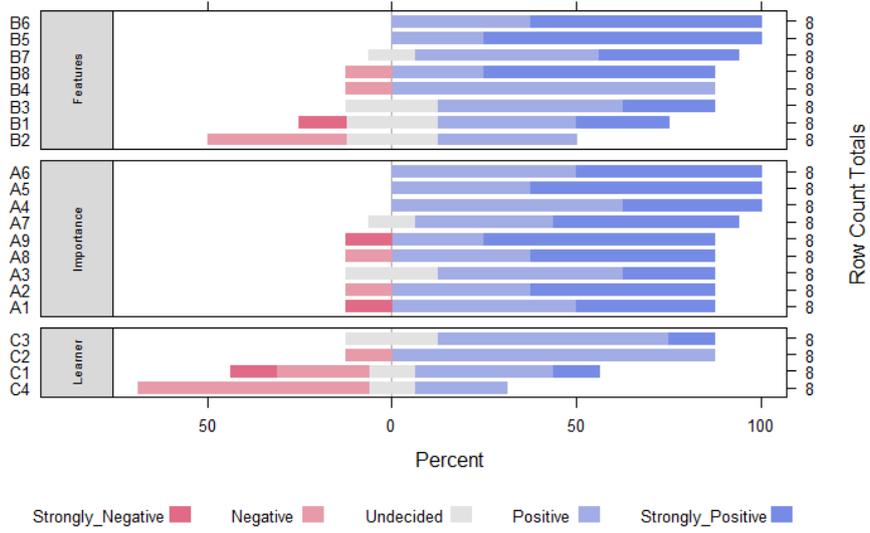

**Fig 5** IDS evaluation – Arts

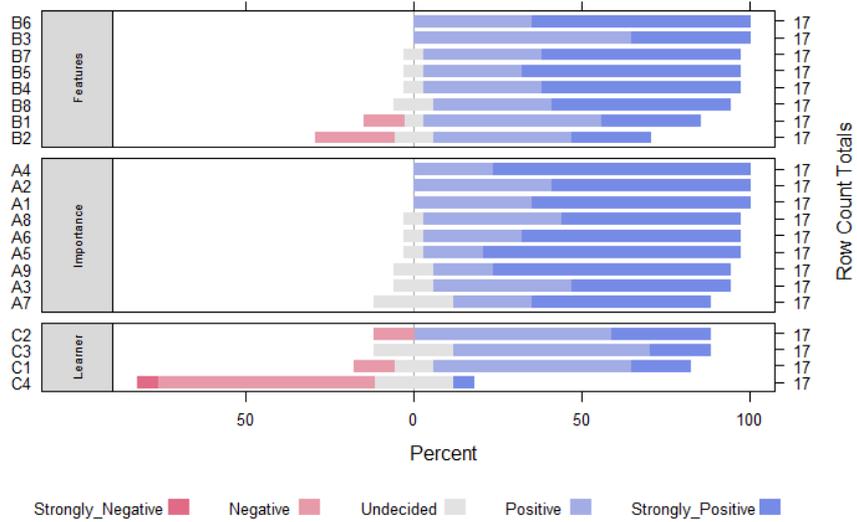

**Fig 6** IDS evaluation – Engineering & Math

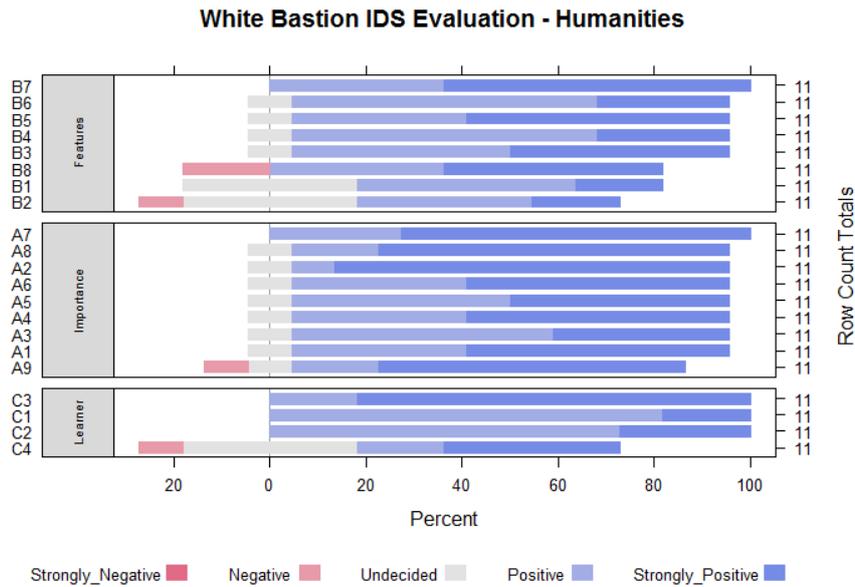

**Fig 7** IDS evaluation – Humanities

The questionnaire included additional set of 4 items addressing the opposition between the artistic liberties and fidelity to technical constraints and historical facts. We included only four questions and put them in the sequence reflecting the level of non-fidelity:

D1.   "The costumes are not genuine as in the age presented in the story."
D2.   "The food or tools used in the story were not known in the age presented in the story."
D3.   "Animations do not comply with technical restrictions."
D4.   "Some facts presented are not baked by the historic evidence."

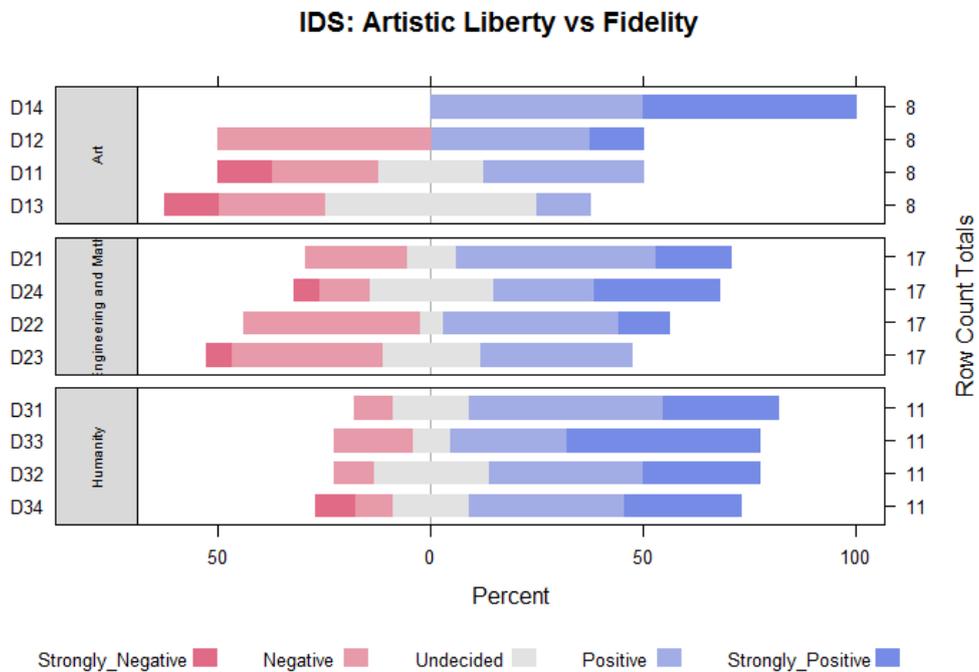

**Fig 8** IDS challenge: Artistic Liberty vs Fidelity

The sample of users involved in this pilot evaluation displayed foreseen patterns, specifically that historians (Humanities) and engineers highlight the importance of fidelity over artistic liberty. Not foreseen is the strict adherence to facts within artist population.

V. CONCLUSION

In this paper we presented research undertaken by an interdisciplinary team of experts from computer science, visual arts, film directing, literature, psychology, communicology and human computer interaction, aimed to develop a new method of interactive digital storytelling for cultural heritage presentation: hyper-storytelling. The experts offered their insights in concept, structure and elements which an IDS presentation needs to contain in order to achieve maximum level of user immersion and offer the most satisfactory edutainment experience. This research shows the need to view IDS as editorial content assembled according to rules of multiple professions, each contributing in its own way to the common product.

In order to include the IDS content as a part of standard educational process, and bring it in the classroom, some additional exploration of user expectations were measured and analyzed. Future work of this team will be performed according to the established guidelines. IDS application intended for educational purposes will be further evaluated with both edutainment levels and effectiveness measured. According to results of these more detailed studies the new IDS concept will be adjusted and finalized.

REFERENCES

Alexander, B., *The New Digital Storytelling: Creating Narratives with New Media*. Westport, CT, USA: Praeger Publishers, 2011.

Boskovic, D. et al, "Measuring immersion and edutainment in multimedia cultural heritage applications" in *2017 XXVI International Conference on Information, Communication and Automation Technologies (ICAT)*, Sarajevo, 2017, pp. 1-6.

Campbell, J., *The Hero with a thousand faces*, first edition. 1949.

Chalmers, P. A., "The role of cognitive theory in human–computer interface," *Computers in Human Behavior*, vol. 19, no. 5, 2003, pp. 593–607.

Coomey, M., and J. Stephenson, (2001). "Online learning: it is all about dialogue, involvement, support and control – According to the research." *Teach. Learn. Online Pedagogies New Technol.* 2001, pp. 37–52.

De Freitas, S. and F. Liarokapis, "Serious Games: a new paradigm for education?" In *Serious Games and Edutainment Applications*. 2011, pp. 9-23.

Denard, H. "A new introduction to the London Charter," in A. Bentkowska-Kafel, D. Baker, and H. Denard, (eds.) *Paradata and Transparency in Virtual Heritage Digital Research in the Arts and Humanities Series*, Ashgate, 2012, pp. 57–71.

DeStefano, D. and J.-A. LeFevre, "Cognitive load in hypertext reading: A review," *Computers in Human Behavior*, vol. 23, no. 3, 2007, pp. 1616–1641.

Heiberger, R.M., and N.B. Robbins, "Design of diverging stacked bar charts for Likert scales and other applications," *Journal of Statistical Software*, vol. 57, no. 5, 2014, pp. 1-32.

Hollender, N. et al., "Review: Integrating cognitive load theory and concepts of human-computer interaction," *Computers in Human Behavior*, vol. 26, no. 6, 2010, pp. 1278–1288.

Kerlow, I. W., *The Art of 3-D Computer Animation and Effects*. John Wiley and Sons, 2009.

Lazar, J., Feng, J.H., and H. Hochheiser, *Research Methods in Human-Computer Interaction*, Wiley Publishing, 2010

Liarokapis, F. et al., "Multimodal Serious Games Technologies for Cultural Heritage" in *Mixed Reality and Gamification for Cultural Heritage,* Cham, Springer International Publishing, 2017, pp. 371-392.

Miller, C., *Digital Storytelling: A Creator's Guide to Interactive Entertainment*. Focal Press, 2004.

Nielsen, J., *10 usability heuristics for user interface design*. Nielsen Norman Group, 1995.

Pagano, A., Armone, G., and E. D. Sanctis, "Virtual museums and audience studies: the case of 'Keys to Rome' exhibition," in *2015 Digital Heritage*, vol. 1, 2015, pp. 373–376.

Philpin-Briscoe, O. et al., "A Serious Game for Understanding Ancient Seafaring in the Mediterranean Sea," *Proceedings of 9th International Conference on Virtual Worlds and Games for Serious Applications VS-Games*, Athens, Greece, 2017, pp. 1-7.

Pietroni, E., Pagano, A., and C. Rufa, "The Etruscanning project: Gesture based interaction and user experience in the virtual reconstruction of the Regolini-Galassi tomb," in *2013 Digital Heritage International Congress (DigitalHeritage)*, vol. 2, 2013, pp. 653–660.

Quintilian, in L. Honeycutt, (ed.), *Institutes of Oratory*, 2006.

Rizvic, S. et al., "4D Virtual Reconstruction of White Bastion Fortress," in C. E. Catalano and L. D. Luca, (eds.), *Eurographics Workshop on Graphics and Cultural Heritage*, The Eurographics Association, 2016.

Schoenau-Fog, H., *Adaptive Storyworlds*. Springer International Publishing, 2015, pp. 58–65.

Sweller, J., Ayres, P., and S. Kalyuga, *Adaptive Storyworlds*. New York: Springer, 2011.

Wiberg, C. and K. Jegers, "Satisfaction and learnability in edutainment: a usability study of the knowledge game 'Laser Challenge' at the Nobel e-museum," *Proceedings of HCI International–10th International Conference on Human Computer Interaction*, Crete, Greece, 2003, pp. 1096-1102.